\documentclass[11pt,a4paper]{article}
\usepackage{jinstpub}
\usepackage{natbib}
\usepackage{amsthm}
\usepackage{graphicx}
\usepackage{xcolor}
\usepackage{subcaption}
\usepackage{indentfirst}
\usepackage{float}

\title{The effects of a passive Bi-Polar Grid (BPG) on \\ Ion Back-Flow (IBF) and Resolution}
\author[a]{V.~Zakharov,}
\author[a]{E.~Shulga}
\author[a]{P.~Garg}
\author[a]{T.~Hemmick}
\author[b]{A.~Milov}

\affiliation[a]{Physics \& Astronomy, Stony Brook University (SBU),\\
100 Nicolls Rd., Stony Brook, United States}
\affiliation[b]{Physics, Weizmann Institute of Science (WIS),\\
234 Herzl St., Rehovot , Israel}

\emailAdd{vladislav.zakharov@stonybrook.edu}

\arxivnumber{}

\abstract{
Time Projection Chambers (TPC)s are excellent tracking detectors for high multiplicity events and can intrinsically be high-rate, but are limited by the ions created in their avalanche stage. GEMs and Micromegas can reduce IBF through their geometry and $\Vec{E}$-field ratios, but these can lead to gain fluctuations and still leave IBF as the dominant source of space charge. An active BPG can block all IBF ions, but their slow drift speed creates too much dead time. A passive BPG will overcome this limitation by using an external $\Vec{B}$-field to allow the electrons to pass through while still blocking all ions. Since the grid changes the electron's trajectory, a loss of resolution will occur. The trajectory is shifted symmetrically along the wires so the wire alignment with respect to the detection pads is a specific question not studied before. We present completed IBF analysis from data collected at Weizmann Institute of Science (WIS), along with an intro to our test on wire resolution.
}


\notoc
\begin{document}
\maketitle

\section{Gating Grids}
\begin{figure}[H]
    \centering
    \captionsetup{justification=centering}
    \begin{subfigure}{0.3\textwidth}
        \includegraphics[width=\textwidth]{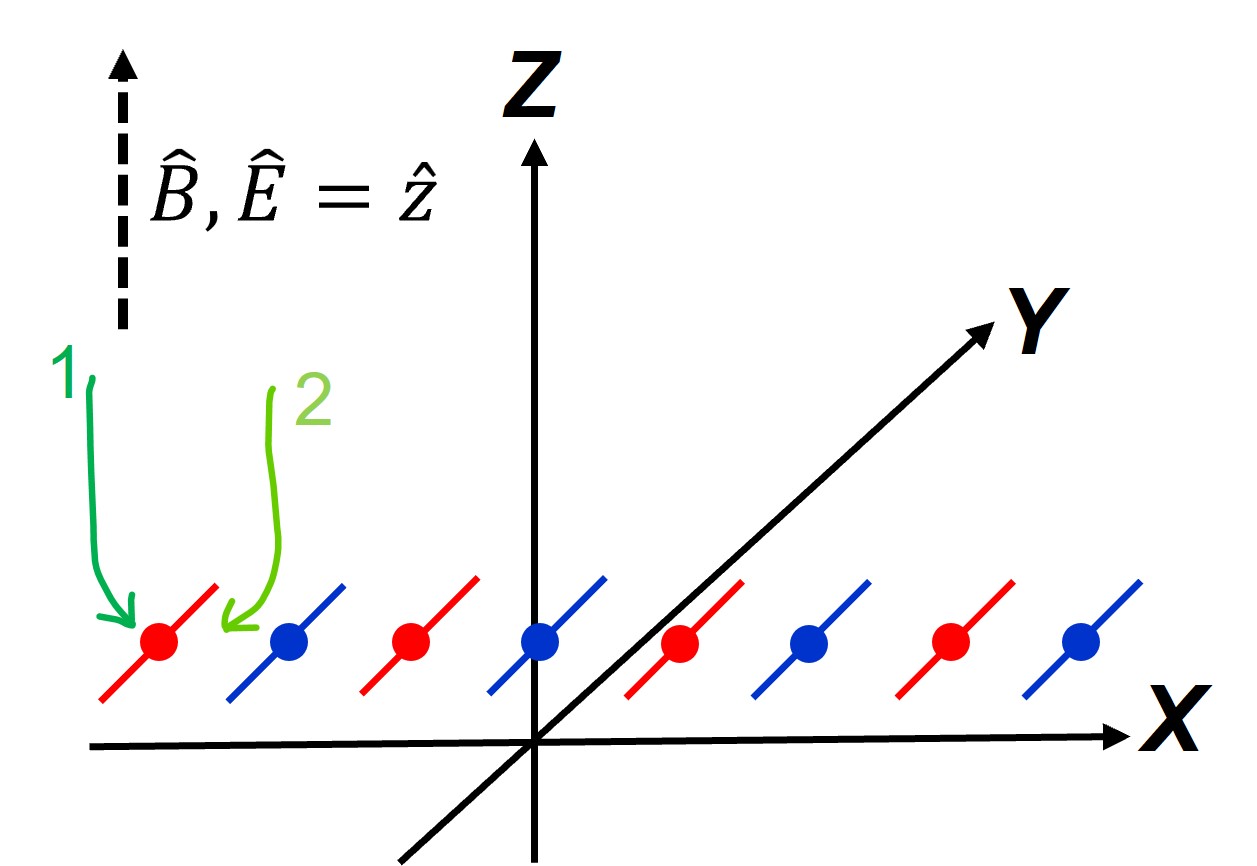}
        %\caption{An illustration of 2 different electrons drifting down from opposite sides of a positive BPG wire}
        \label{my drawn coordinates} 
    \end{subfigure}
    \qquad
    \begin{subfigure}{0.2\textwidth}
        \includegraphics[width=\textwidth]{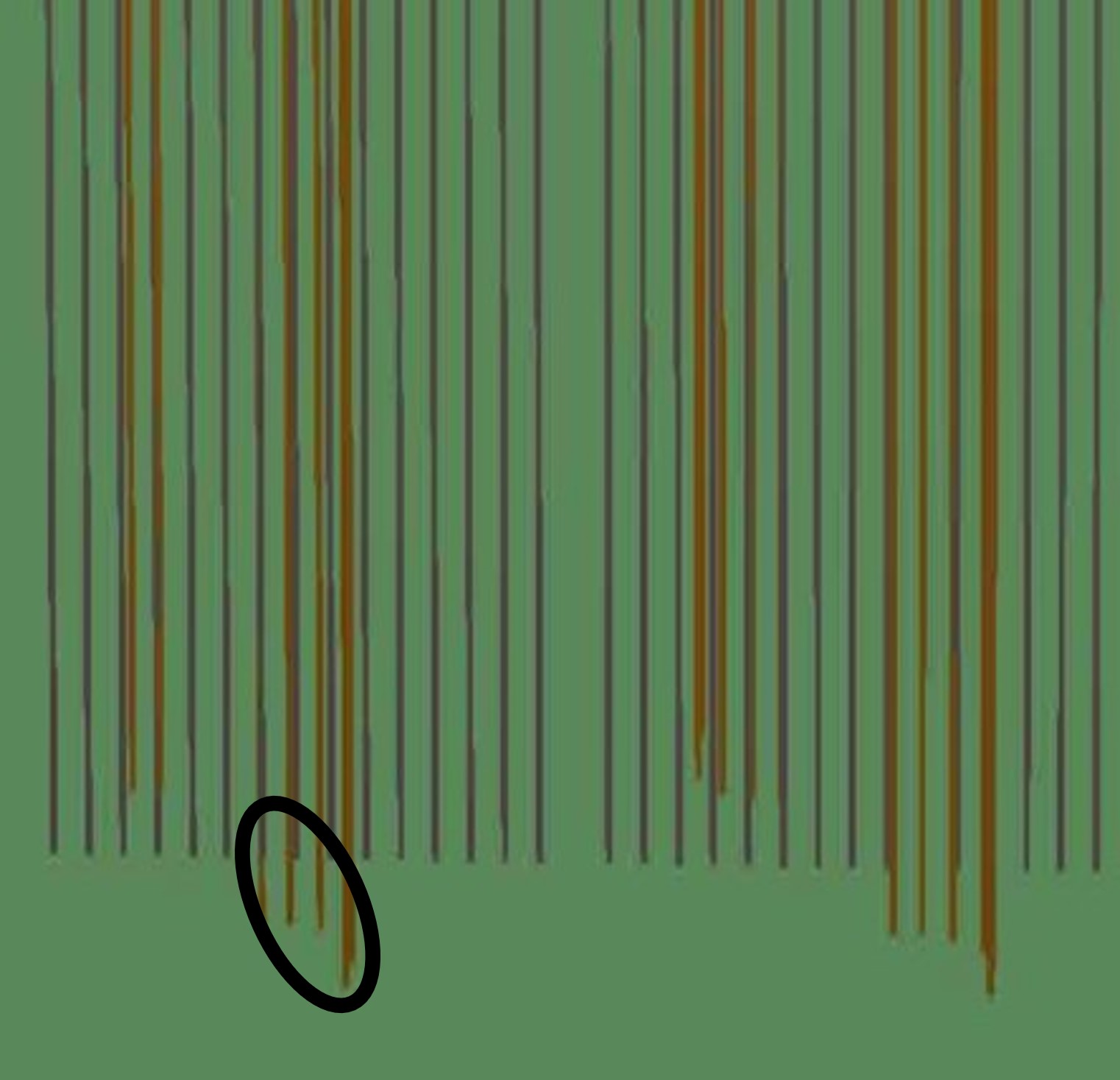}
        %\caption{\cite{Zakharov:2019qmq} Simulation close-up showing electrons in orange lines making it to the green plane past the BPG. Blue lines show ion paths}
        \label{out of plane zoom} 
    \end{subfigure}
    \caption[Electrons passing through a passive BPG]
    {Left: Illustration of electrons drifting from opposite sides of a positive BPG wire. Right: \cite{Zakharov:2019qmq} Simulation close-up showing electrons in orange lines making it to the green detection plane past the BPG. Blue lines show ion paths. The black oval indicates the electron distortions.}
    \label{electrons passing grid}
\end{figure}

A passive BPG with $(V_g \pm \Delta g)$ can provide $\Vec{E}$-field ratios without creating fluctuations and would eliminate dead time by constantly being opaque to ions while exploiting an external magnetic $\Vec{B}$-field to remain transparent to electrons. 
\begin{equation}
    \label{langevin}
    m\frac{d\vec{v}}{dt} = q\vec{E} +q(\vec{v} \times \vec{B}) -\kappa\vec{v}
\end{equation}
Charges follow the Langevin equation and lighter electrons will experience a continuous $(\vec{v} \times \vec{B})$ ``push" allowing them to pass between the wires. Electrons coming from above the wire will get their 1\textsuperscript{st} push along the length of the wire: $\hat{F_1} = \pm \hat{x} \times \hat{z} = \mp \hat{y}$. Then they get a 2\textsuperscript{nd} weaker push which allows them to get past the wires: $\hat{F_2} = \mp \hat{y} \times \hat{z} = \mp \hat{x}$. Fig.\ \ref{electrons passing grid} shows these effects along with the distortion both pinched and shifted in their y-position. Acceptable electron transparency with 0\% ion transparency has been demonstrated by previous groups \cite{Amendolia:1985dk}. 

The general equation for the charge velocity in a gas as a function of $\vec{E}$ and $\vec{B}$ fields is given by
\begin{equation}
    \label{charge mobility Allis}
     \vec{v} = \frac{\mu}{1+\omega^2\tau^2} \left(\vec{E} 
     +\frac{\vec{E} \times \vec{B}}{|\vec{B}|}\omega \tau
     +\dfrac{(\vec{E} \cdot \vec{B})\vec{B}}{B^2}\omega^2\tau^2 \right)
\end{equation}
where $\mu=e\tau/m$ is the charge mobility with $\tau$ as the average time between collisions. $\omega =eB/m$ is the cyclotron frequency. Defining angle $\alpha$ between $\vec{E}$ and the positive x-direction gives $\vec{E}=(E\cos{\alpha}, 0, E\sin{\alpha})$. Together with $\vec{B}=(0, 0, B)$ leads to 
\begin{equation}
     v_x = c(E \cos{\alpha}) \qquad
     v_y = c(E \omega \tau \cos{\alpha}) \qquad
     v_z = \mu E \sin{\alpha}
\end{equation}
If the BPG wires aren't uniformly spaced along their tensioned direction an $E_y$ component will be present and the charge velocity equations become 
\begin{equation}
    \label{non-parallel wires}
    v_x = c(E_y \omega\tau +E_x) \qquad
    v_y = c(-E_x \omega\tau +E_y) \qquad
    v_z = c(E_y \omega^2 \tau^2 +E_z)
\end{equation}
where $c=\dfrac{\mu}{1+ \omega^2 \tau^2}$. 
In both cases, the velocities of main interest for ion blocking and electron transparency $v_x$ and $v_z$ are strongly dependent on $\omega \tau$. For electrons $\omega \tau \gg 1$ and the velocities will follow the $\vec{B}$-field and ignore the local $\Vec{E}$-field of the BPG. For ions the converse is true. 

\section{Weizmann IBF Measurements}
\begin{figure}[H]
    \centering
    \includegraphics[width=0.7\textwidth]{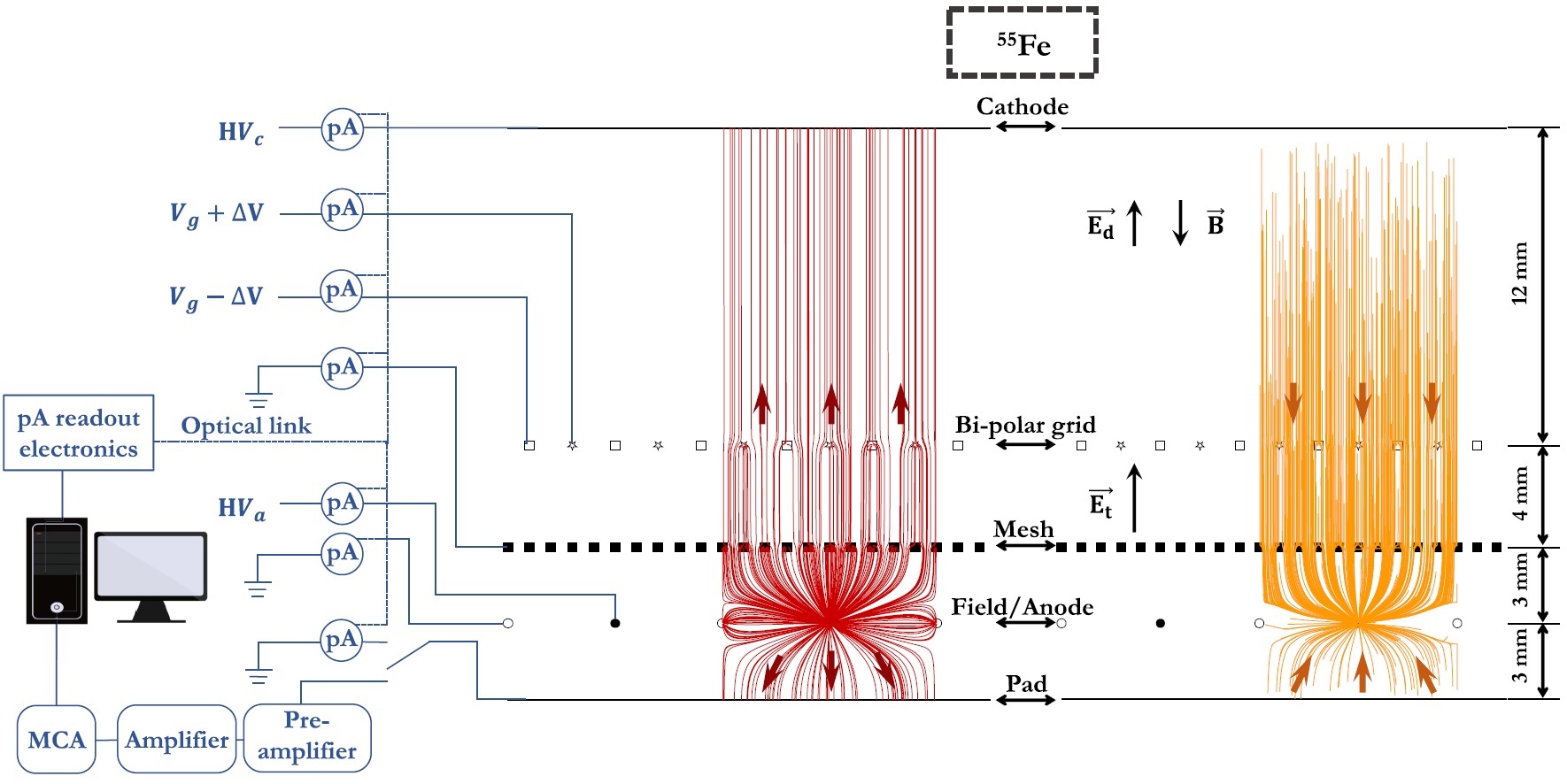}
    \caption{Illustration showing all elements of the WIS detector. Orange and red lines are electrons and ions respectively.}
    \label{WIS setup} 
\end{figure}

Ion and electron transparencies, as a function of $\vec{B}$-field and $\Delta V$ volts on a linear BPG, were measured at Weizmann. A detector with MWPC gain wires was constructed. A strong \textsuperscript{55}Fe X-Ray source was used and picoAmmeters were implemented to measure the small changes in current as the grid was slowly closed to ions. A mesh was used to help create a uniform drift field, as seen in Fig.\ \ref{WIS setup}, and to create a region where the $\Vec{E}$-field above-to-below ratios can be better exploited. 

\begin{figure}[H]
    \centering
    \begin{subfigure}{0.4\textwidth}
        \includegraphics[width=\textwidth]{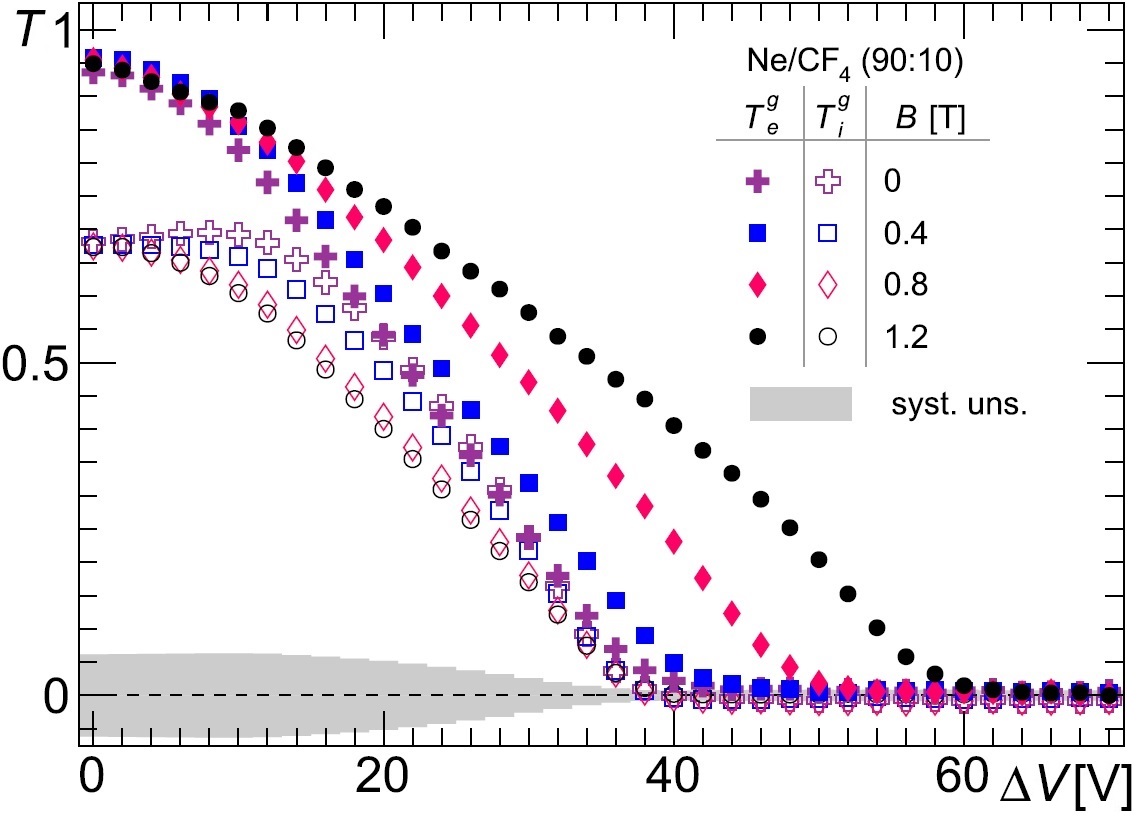}
        \caption{Ion and electron grid transparencies}
        \label{IBF data}
    \end{subfigure}
    \qquad \qquad
    \begin{subfigure}{0.4\textwidth}
        \includegraphics[width=\textwidth]{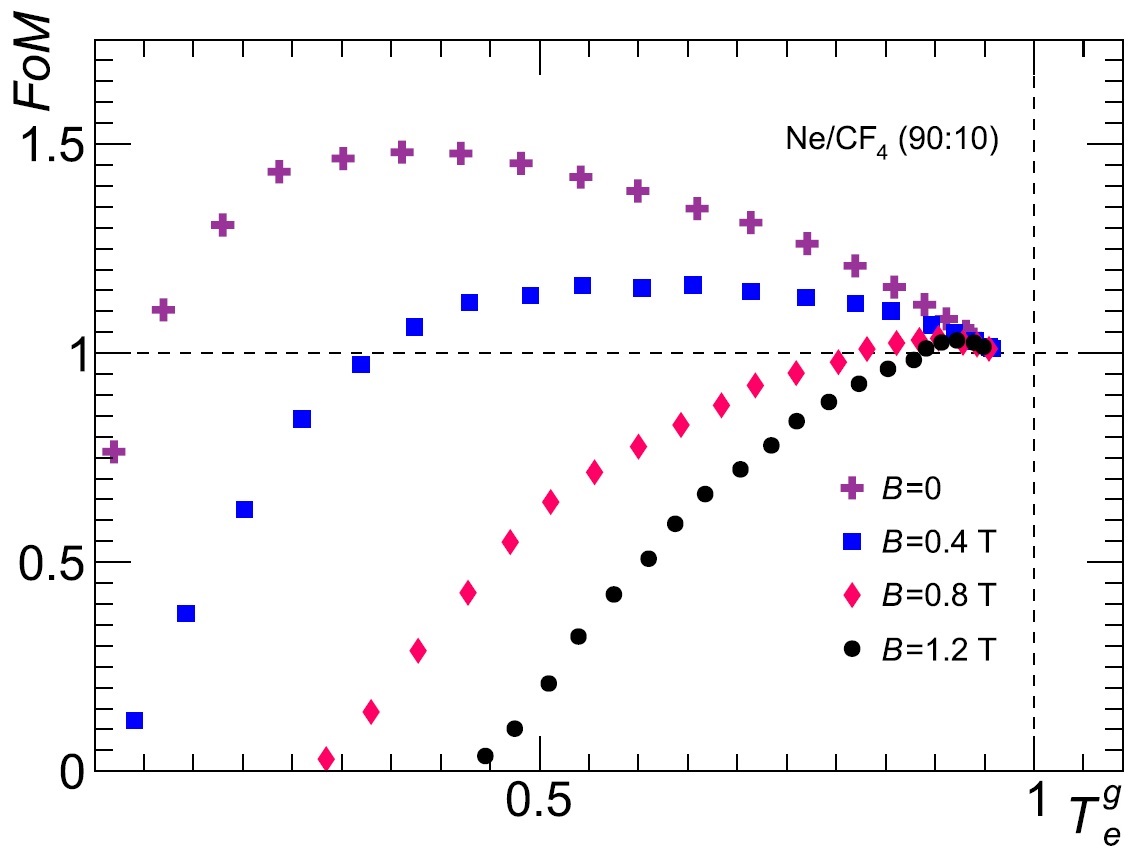}
        \caption{FoM demonstrating BPG effectiveness}
        \label{FOM}
    \end{subfigure}
    \captionsetup{justification=centering}
    \caption[IBF results]{Some IBF results from Weizmann. Full results: \cite{Zhenia:9279256}}
    \label{transparency and FoM}
\end{figure}

Field ratios were varied measuring the electron and ion transparencies from the BPG and mesh $T^{g,m}_{e,i}$. Increased grid ratios meant decreased mesh ratios pulling more ions through it. $T_e^g <100\%$ requires an increased gain, which creates more IBF. A Figure of Merit (FoM) was constructed to quantify the effectiveness of the BPG \cite{Zhenia:9279256}. 1\textsuperscript{st} term comes from $\Vec{E}$ ratios, where higher ratios can extract more ions from the gain region. 2\textsuperscript{nd} term compensates for loss of primaries creating more IBF ions. Fig.\ \ref{transparency and FoM} shows an example for transparencies and FoM.

\begin{equation}
    FoM(\omega,\Delta V)= \frac{T^m_i(\omega,0)}{T^m_i(1,0)} 
                    \cdot \frac{T^g_i(\omega,\Delta V)}{T^g_e(\omega,\Delta V}
    \qquad \omega=E_t/E_d
\end{equation}

\section{BPG Resolution Distortion}
From the equations of motion, and as seen in Fig.\ \ref{electrons passing grid}, the electrons fall "out of plane", and alternating wires "pinch" them together, hurting resolution. The error is cyclical with the same periodicity as the wires. These recurring shifts are known as Differential Non-Linearity (DNL), and can be corrected. Segmented pads are used to spread out the electron cloud charge collection with finer detail in order to get a more precise track position measurement. 

\begin{figure}[H]
    \centering
    \begin{subfigure}{0.4\textwidth}
        \includegraphics[width=\textwidth]{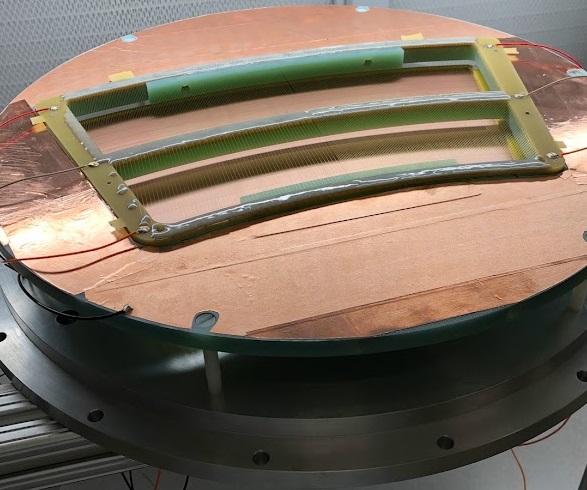}
        \caption{Wires of four different configurations placed above a quad-GEM stack}
        \label{BPG with module}
    \end{subfigure}
    \hfill
        \begin{subfigure}{0.45\textwidth}
        \includegraphics[width=\textwidth]{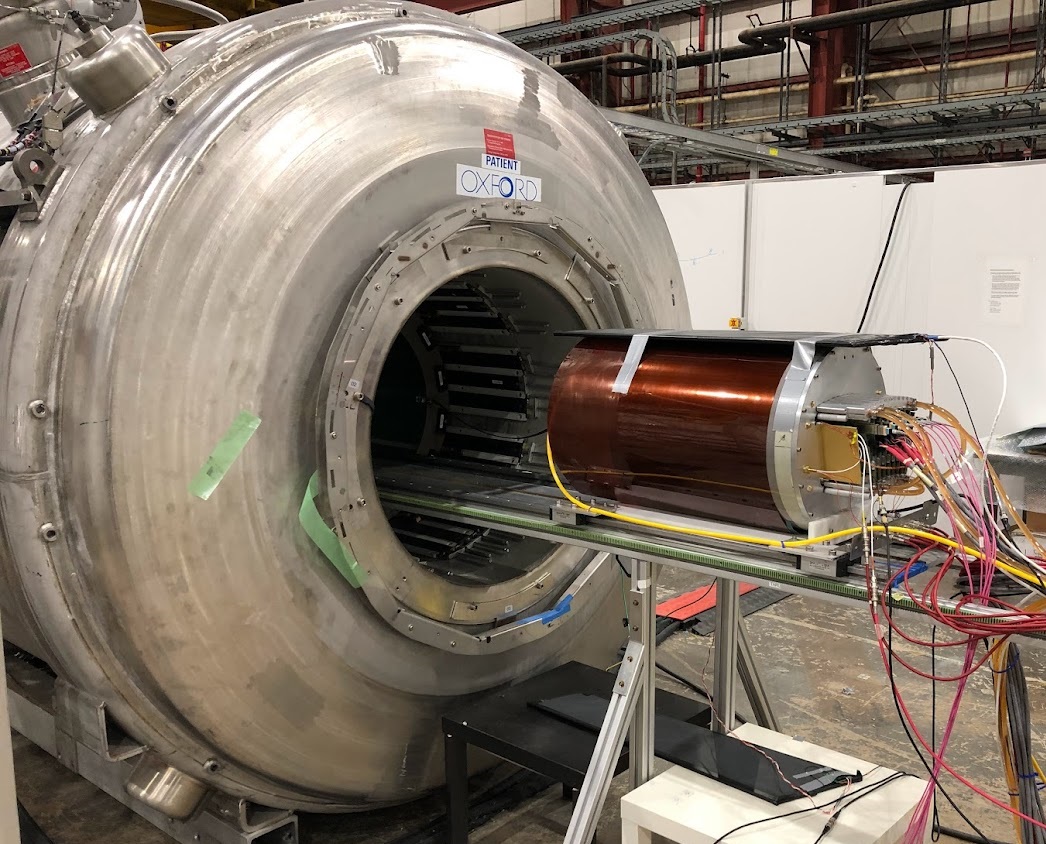}
        \caption{Prototype TPC about to enter $\Vec{B}=3T$ at ANL}
        \label{TPC at ANL}
    \end{subfigure}
    \captionsetup{justification=centering}
    \caption[Argonne trip]{Recording position resolution data using cosmic rays passing through a TPC.}
    \label{ANL test}
\end{figure}

Zig-zag pads \cite{Azmoun:8379440}, which themselves have DNL-corrected error due to their shape, are used. From the grid's symmetric shift, perhaps BPG's distortions will average out over the full particle track, regardless of the wire's position with respect to the radial zig-zags. However, if the grid is aligned with the pads then their independent DNLs might combine for better resolution. A BPG with 2 linear and 2 radial wire configurations is placed above the pad-plane and goes inside a prototype TPC. We've taken it to Argonne National Lab (ANL) to test this hypothesis (Fig.\ \ref{ANL test}). 

\section{Conclusion}
The passive BPG source test demonstrates ion blocking with high electron transparency. The ANL test showed excellent resolution results and a paper with the full test description and analysis will soon be published.

\bibliographystyle{JHEP}
\bibliography{references}

\end{document}